\documentclass[conference]{IEEEtran}

\usepackage{cite}
\usepackage{amsmath,amssymb,amsfonts}
\usepackage{algorithmic}
\usepackage{graphicx}
\usepackage{textcomp}
\usepackage{xcolor}
\usepackage{blkarray, bigstrut}
\usepackage{bm}
\usepackage{nicematrix}
\usepackage{enumerate}

\def\BibTeX{{\rm B\kern-.05em{\sc i\kern-.025em b}\kern-.08em
    T\kern-.1667em\lower.7ex\hbox{E}\kern-.125emX}}

\begin{document}
\bstctlcite{MyBSTcontrol}

\title{A Merge/Split Algorithm for Multitarget Tracking Using Generalized 
Labeled Multi-Bernoulli Filters}

\author{\IEEEauthorblockN{Lingji Chen}
\IEEEauthorblockA{\textit{Motional} \\ Boston, USA \\
chen-lingji@ieee.org}}

\maketitle

\begin{abstract}
The class of Labeled Random Finite Set filters known as the delta-Generalized 
Labeled Multi-Bernoulli (dGLMB) filter represents the filtering density as a 
set of weighted hypotheses, with each hypothesis consisting of a set of labeled 
tracks, which are in turn pairs of a track label and a track kinematic density. Upon 
update with a batch of measurements, each hypothesis gives rise to many child 
hypotheses, and therefore truncation has to be 
performed for any practical application. Finite compute budget can lead to degeneracy that drops tracks. To mitigate, we adopt a 
factored filtering density through the use of a novel Merge/Split algorithm. Merging has long been established in the literature; our splitting algorithm is enabled by an efficient and effective marginalization scheme, through indexing a kinematic density by the measurement IDs (in a moving window) that have been used in its update. This allows us to determine when independence can be considered to hold approximately for a given tolerance, so that the ``resolution'' of tracking is adaptively chosen, from a single factor (dGLMB), to all-singleton factors (Labeled Multi-Bernoulli, LMB), and anywhere in between.  
\end{abstract}


\IEEEpeerreviewmaketitle

\section{Introduction}

Multitarget 
tracking is a challenging problem that can be solved in the framework of Joint 
Probabilistic Data Association (JPDA), Multiple Hypothesis Tracking 
(MHT), and Random Finite Set (RFS); see a recent survey paper \cite{VoBN15_mtt} 
with a comprehensive list of references. A class of Labeled RFS filters known 
as the $\delta$-Generalized Labeled Multi-Bernoulli ($\delta$-GLMB) filter has 
been shown to provide a ``closed form'' solution to such tracking problems 
\cite{VoBT13,VoBN14}, and many successful applications have been reported in 
the literature. As has been shown in \cite{VoB17} (and stated more explicitly 
in \cite{ChenL18_code,ChenL18}), a $\delta$-GLMB filter represents the filtering density as 
a set of weighted hypotheses, with each hypothesis consisting of labeled 
tracks\footnote{or more precisely, labeled track {\em points} at the current 
time. In this paper we use ``track'' and ``track point'' interchangeably, and will state it explicitly when we mean a track {\em history} over time.}, which are in turn pairs of a track 
label and a track kinematic density.  Upon update with a batch of 
measurements, each hypothesis gives rise to many child hypotheses \cite{VoB17}, 
and therefore for any practical application, truncation has to be performed in 
order to fit a given compute budget, expressed for example as the maximum 
number $K$ of hypotheses held simultaneously in memory.

This filter in a sense can be viewed as a modern day treatment of the Hypothesis Oriented MHT (HO-MHT) first proposed in \cite{ReidD79}. It has been known for decades that clustering of tracks and measurements should be exploited so that a single, global tracking problem can be solved as parallel, local tracking sub-problems, with less computational cost. More concretely, suppose that a platform carries two sensors, one looking 
eastward and the other looking westward. If each sensor runs a tracker on its own data 
with say 10 hypotheses, then we need only $10+10=20$ hypotheses, under the assumption of independence, to characterize the whole scene. However, if we 
use a single, global tracker that ingests measurements from both 
sensors without consideration of independence, then we would need $10 \times 10 
= 100$ hypotheses, enumerating the combinations of local hypotheses on the east with local hypotheses on the west.

This is not merely an issue of efficiency, because of the following competing characteristics of a hypothesis-enumerating tracker that is practically viable:
\begin{itemize}
 \item We typically want a track confirmation stage, i.e., we want a hypothesis for a new track to accumulate evidence and increase weight as more frames of measurements are received.
 \item We are invariably limited by a compute budget; we have to truncate and keep only a subset of the hypotheses in memory.
 \item We typically rank the hypotheses by their weights and truncate accordingly. 
\end{itemize}
What this means is that under a given budget, degeneracy may occur in the following sense: When a new target appears on the east, to continue with the above example, the candidate hypothesis $h$ will have a small weight, and if we look at the ranking of the global hypotheses (in the form of east-west combination), we may find that those above $h$ are simply enumerating minute variations of measurement to track associations due to the west side of sensor data. If $h$ happens to be cut off by the budget, then the measurement of this track in the next frame will suffer from the same fate, leading to the track being missed altogether. 

One way to mitigate such degeneracy is to use a factored representation, so that the budget for the east is not tied to the budget for the west, and a nascent track hypothesis has a chance to be kept and to accumulate more evidence. A factored representation requires the operations of merging and splitting: When previously independent clusters become coupled by the newly arrived measurements, the corresponding factors need to be merged one way or another. On the other hand, resolved uncertainty leads to pruning and possibly independence, and splitting may be performed to prevent the case of everything coalescing into one gigantic factor. Merging is well understood, while splitting has been very challenging \cite{RoyJ97_efficient,BeardM20_solution}. In this paper we rigorously define these operations, that are implementation-friendly, and offer two new ideas: One is to control the tolerance of when independence can be declared to approximately hold, the other is an extremely efficient way of performing marginalization of a $\delta$-GLMB density by indexing a track's kinematic density with a moving window of the incorporated measurement IDs.

The paper is a substantial revision of an earlier draft, and is organized as follows. In Section~\ref{sec:factor} we present the factored form  of $\delta$-GLMB and define merging and the desired condition for splitting. In Section~\ref{sec:indexing} we discretize the problem by indexing a kinematic density with the measurement IDs that have been incorporated into it, in a moving window, so that two densities can be easily identified as being almost the same from having been updated with the same set of recent measurements. Section~\ref{sec:ms} presents the operational details of the Merge/Split algorithm, with an illustration based on simulated data given in Section~\ref{sec:simu}. The connections between the proposed algorithm and the approaches in the literature are discussed in Section~\ref{sec:related}, with conclusions drawn in Section~\ref{sec:con}. A specialized $K$-min-sum algorithm for the selection of the best $K$ merged hypotheses is presented in the Appendix with a proof of correctness.

%\section{Selecting the best $K$ hypotheses} \label{sec:hypo}
\section{The factored $\delta$-GLMB} \label{sec:factor}
\subsection{Paremeterized representation}
The posterior we are interested in is a $\delta$-GLMB distribution over the labeled random finite set $\bm{X}$, defined as
\begin{equation} 
 \bm{\pi}(\bm{X}) = \Delta(\bm{X}) \sum_{i=1}^{N} w^{(i)} \delta_{I^{(i)}}\left[\mathcal{L}(\bm{X})\right]  \prod_{\bm{x} \in \bm{X}} p^{(i)}(\bm{x}),
\end{equation}
where $\Delta(\bm{X})$ ensures that $\bm{X}$ is well formed, $w^{(i)}$ is the weight of the $i$-th hypothesis out of a total of $N$, $\delta_{I^{(i)}}\left[\mathcal{L}(\bm{X})\right]$ states that its label set is $I^{(i)}$, and $p^{(i)}(\cdot)$ denotes the kinematic density of each labeled track in this hypothesis. For details, readers are referred to \cite{VoB17}.

The above expression provides a ``point-wise evaluation'' of the probability density on $\bm{X}$. Just like in the case of a Gaussian random variable, where it is often convenient to use the parameterized representation of the mean and the covariance in a discourse, without resorting to the evaluation formula, we will, for the rest of this paper, use only the parameterized representation of $\bm{\pi}(\bm{X})$ given by 
\begin{equation} \label{eqn:pi}
 \bm{\pi}(\bm{X}): \left\{ \left(w^{(i)}, h^{(i)}\right) \right\}_{i=1}^N \triangleq \left\{ \left(w^{(i)}, \left\{ \left(\ell, p_\ell(\cdot) \right) \right\}_{\ell \in I^{(i)}} \right) \right\}_{i=1}^N.
\end{equation}
In other words, a $\delta$-GLMB is a set of hypotheses with weights, with each hypothesis being a set of tracks, each in turn having a label and a probability density function. Sometimes we use the notations $h \in \bm{\pi}$ to range over the hypotheses, and $\ell \in h$ or $p_\ell \in h$ to range over all the tracks in a hypothesis $h$. The continuous, infinite dimensional $p_\ell(\cdot)$ is treated here as being ``atomic;''\footnote{which is how it is treated when we say that GLMB is ``closed'' under Bayesian operations.} for any practical implementation it has to have a finite representation, be it a Gaussian, a Gaussian mixture, or a set of particles.  

\subsection{Localized tracks}
In general, the probability density $p_\ell(\cdot)$ can be multimodal. This is often the result of a multimodal measurement likelihood function. For example, a radar may report a range rate that is the true value modulo some span: A target may be reported as incoming, at a radial speed of $-v$, but it can actually be outgoing, at a radial speed of $u$, where $u + v = D$, a radar parameter. We contend that the way we think about a multimodal target density is actually through multiple hypotheses: Either the target is doing this, or it is doing that. Therefore, when there is a discrete ``mode number'' for the measurement, we will extend the hypothesis space along this dimension too, and split the corresponding hypothesis so that each has a unimodal probability density function (pdf). More specifically, if the pdf is bimodal,
\[
 p_{\ell_1}(\cdot) = \alpha \, p_{\ell_1}^{(\alpha)} (\cdot) + \beta \, p_{\ell_1}^{(\beta)} (\cdot),
\]
then the hypothesis
\[
 (w_i, h_i) = \left(w_i, \{(\ell_1, p_1(\cdot)), \_\} \right),
\]
where the underscore ``\_'' represents ``the rest,'' is replaced by two hypotheses
\[
 \left(\alpha \, w_i, h_i^{(a)} \right) \triangleq \left(\alpha \, w_i, \{(\ell_1, p_1^{(\alpha)} (\cdot)), \_\} \right),
 \]
 and
 \[
 \left(\beta \, w_i, h_i^{(\beta)} \right)  \triangleq \left(\beta \, w_i, \{(\ell_1, p_1^{(\beta)} (\cdot)), \_\} \right).
 \]
This process may have to be repeated until all pdf's are unimodal.

In tracking, coarse gating is routinely used to eliminate those data associations that are practically impossible. The gating operation can be carried out without actually performing the prediction, e.g. by using the maximum speed of a target and the elapsed time. This leads us to consider a track's pdf as having a {\em finite} support. Let the measurement function for the frame of sensor data be defined by $g(\cdot)$, and let $\gamma_\ell(z, p_\ell)$ be the boolean coarse gating function for Track $\ell$ such that it returns $1$ if and only if we consider the association between $z$ and this track possible. For example, when $p_\ell$ is Gaussian, we can define a ``rectangle'' that is big enough to account for both the ``three sigma'' of $p_\ell$ and the uncertainty in $z$. Then the track's finite support is defined by
\[
 \mathcal{D}(p_\ell) = \{x \in \mathcal{X}: \gamma_\ell(g(x), p_\ell)=1 \}.
\]
Then it follows that a hypothesis has a finite support too, given by 
\[
 \mathcal{D}(h) = \bigcup_{p_\ell \in h} \mathcal{D}(p_\ell),
\]
and a $\delta$-GLMB, given by (\ref{eqn:pi}), has a finite support as well, given by 
\[
 \mathcal{D}(\bm{\pi}) = \bigcup_{h \in \bm{\pi}} \mathcal{D}(h).
\]
For convenience we also define its label set
\[
 \mathcal{L}(\bm{\pi}) = \bigcup_{h \in \bm{\pi}} \mathcal{L}(h) = \bigcup_{i=1}^N I^{(i)}.
\]

\subsection{Factored representation, merging, and the challenge of splitting}
Thus, we can view $\bm{\pi}$ in two ways: One is that it describes the entire space, leaving no unspecified targets. The other is that it describes only a factor, restricted to the space of the Cartesian product of $\mathcal{D}(\bm{\pi})$ and $\mathcal{L}(\bm{\pi})$, and says nothing about the rest of the space.

With the second perspective, we can define the product of two factors, which is the merging of hypotheses. Let $\bm{\pi}_1$ and $\bm{\pi}_2$ be defined in the style of (\ref{eqn:pi}) with $N$ and $M$ hypotheses respectively. If they are disjoint, i.e., 
\begin{equation} 
\mathcal{D}(\bm{\pi}_1) \cap \mathcal{D}(\bm{\pi}_2) = \emptyset, \; 
\mathcal{L}(\bm{\pi}_1) \cap \mathcal{L}(\bm{\pi}_2) = \emptyset,
\end{equation}
then the representation of their product is given by 
\begin{equation} \label{eqn:merging}
\begin{split}
 \bm{\pi}_1 \otimes \bm{\pi}_2  & : \left\{ \left(w^{(ij)}, h^{(ij)}\right)\right\} , i=1, \ldots, N, j=1, \ldots, M  \\
 & \triangleq \left\{ \left(w^{(i)} w^{(j)}, \left\{ \left(\ell, p_\ell(\cdot) \right) \right\}_{\ell \in I^{(i)}} \cup \left\{ \left(s, p_s(\cdot) \right) \right\}_{s \in I^{(j)}}  \right)  \right\}, \\
 & \quad i=1, \ldots, N, j=1, \ldots, M.
\end{split}
 \end{equation}
It can be seen that the product of the factors represents the merging of the hypotheses. 

While merging can be carried out at any time, usually after two or more factors are ``coupled'' by new measurements, the reverse operation of splitting may be possible only under certain conditions. First let us consider a simplified case. Let 
 \[
  \bm{\pi}: \left\{ \left(w^{(t)}, h^{(t)} \right) \right\}_{t=1}^T
 \]
be a set of $T$ weighted hypotheses such that each hypothesis $h^{(t)}$ consists of two subsets, the left $h_L^{(t)}$ and the right $h_R^{(t)}$, with the properties that 
\begin{enumerate}
 \item The cardinalities of the left set and the right set are $N$ and $M$ respectively:
 \begin{equation} \label{eqn:LR}
%\begin{split}  
\left\{h_L^{(t)}\right\}_{t=1}^T = \left\{h_L^{(i)}\right\}_{i=1}^N, \;
 \left\{h_R^{(t)}\right\}_{t=1}^T = \left\{h_R^{(j)}\right\}_{j=1}^M.
%\end{split}
 \end{equation}

  \item $T = N M$. 
 \item The subspaces are disjoint:
 \[
 X_L \triangleq \bigcup_{i=1}^N \mathcal{D}(h_L^{(i)}), \; X_R \triangleq \bigcup_{j=1}^M \mathcal{D}(h_R^{(j)}), \; X_L \cap X_R = \emptyset,
 \]
\[
 I_L \triangleq \bigcup_{i=1}^N \mathcal{L}(h_L^{(i)}), \; I_R \triangleq \bigcup_{j=1}^M \mathcal{L}(h_R^{(j)}), \; I_L \cap I_R = \emptyset.
 \]
\end{enumerate}
Then we can reparameterize $\bm{\pi}$ as
\begin{equation} \label{eqn:LR-2}
 \bm{\pi}: \left\{ \left(w^{(ij)}, \left(h_L^{(i)}, h_R^{(j)}\right) \right) \right\}, i = 1, \ldots, N, j=1, \ldots, M.
\end{equation}

Define a discrete distribution as
\[
 P(i, j) = w^{(ij)}, i = 1, \ldots, N, j=1, \ldots, M.
\]
The marginals are given by 
\[
 w^{(i)} = P(i) = \sum_{j=1}^M P(i, j), \; w^{(j)} = P(j) = \sum_{i=1}^N P(i, j).
\]
If we attempt to assume independence and represent $P(i, j)$ as $P(i) P(j)$, then the reconstruction error is 
\begin{equation} \label{eqn:eps}
 \epsilon = \max_{i,j} |P(i, j) - P(i) P(j)|.
\end{equation}

If independence indeed holds, i.e., $\epsilon = 0$, then we can define two factors, the left and the right, 
\begin{equation} \label{eqn:split}
 \bm{\pi}_L = \left\{ \left(w^{(i)}, h_L^{(i)} \right) \right\}_{i = 1}^N, \;
 \bm{\pi}_R = \left\{ \left(w^{(j)}, h_R^{(j)} \right) \right\}_{j = 1}^M,
\end{equation}
such that 
\[
 \bm{\pi} = \bm{\pi}_L \otimes \bm{\pi}_R.
\]
In practice we rarely get $\epsilon = 0$. But we can have a tolerance for independence, i.e., we require only $\epsilon \ll 1$. However, the bigger challenge is to group hypotheses that are ``similar'' so that $T \approx N M$, because in the extreme case of $T = N = M$, even though we can fill in the missing values with zeros when we define $P(i, j)$, the ``probability table'' would be diagonal after reordering, so that independence would definitely not hold, not even approximately.

\section{Indexing of probability density functions} \label{sec:indexing}
A track's pdf is infinite dimensional. However, we often observe that two tracks are almost the same except that they incorporated slightly different measurements in the long past. This motivates us to adopt the following approach to discretize the pdf's. We identify a track's pdf with the sequence of measurement IDs that have been assigned to the track for its update. Such a sequence increases in 
length as filtering progresses. However, it is often the case with a lot of sensors that  
the effect of old measurements diminishes quickly. In other words, if two tracks 
 have incorporated the same set of measurements for the last, say, 5 updates, 
then their difference is usually negligible. Thus we use a 
fixed-length moving window to keep the most recent $N_w$ measurement IDs as an 
{\em identifier} of the track's pdf. Then we can {\em index} a pdf by the pair {\tt 
(track\_id, density\_id)}, and subsequently index a hypothesis by {\tt 
hypo\_id} which is defined as the sequence of such pairs sorted by {\tt track\_id}. 

This immediately gives us, almost as a byproduct, a convenient way to carry out some hypotheses management: After a measurement update, some old measurement IDs drop out of, while some new measurement IDs come into, the moving window that indexes a track's pdf, such that two previously distinct hypotheses may now be {\em identical}. We keep only one copy from the hypotheses with the same {\tt hypo\_id}, and replace its weight with the sum of all their weights. 

More importantly, when we try to find ``the left set'' and ``the right set'' to split hypotheses, in the sense of Equations (\ref{eqn:LR}) and (\ref{eqn:LR-2}),  we may find many repetitions of the same sub-hypotheses, so that $T \approx N M$. This could lead to a probability table $P(i, j)$ that is denser, thereby is more likely to hold the independence property, within a given tolerance. 

When and how this ``trying'' to split is attempted will be presented in the following sections.

\section{The Merge/Split algorithm} \label{sec:ms}
\subsection{The general $\delta$-GLMB algorithm}
For ease of reference, we summarize the main steps in $\delta$-GLMB filtering, with measurement update  performed jointly with  
motion prediction as is described in \cite{VoB17}. For each hypothesis, there 
is an associated Labeled Multi-Bernoulli (LMB) birth model; by treating birth probability as survival 
probability ``from nothing,'' we can treat both existing tracks and newborn 
track candidates in the same way, and refer to them simply as tracks. 

To determine the most likely ways of associating tracks with measurements, we 
construct a likelihood matrix that has rows for tracks, and three column blocks 
for measurements, missed detections, and deaths respectively. 
This matrix layout is shown here in Table~\ref{tab:layout} for ease of reference; 
it is a simplified version of Figure 1 in \cite{VoB17}, before taking the logarithm of the  
likelihood ratios. 

\begin{table*}
\centering
\begin{blockarray}{cccccccccc}
 ~ & ~ & \mbox{detected} & ~ & ~ & \mbox{missed} & ~ & ~ & \mbox{died} & ~  \\
~ &$z_1$ &  $z_j$ &$z_M$ & $\bar{T}_1$ & $\bar{T}_i$ & $\bar{T}_N$ & 
$\hat{T}_1$ & $\hat{T}_i$ & $\hat{T}_N$\\
\begin{block}{c[ccc|ccc|ccc]}
\bigstrut[t]
$T_1$  & $\star$ & $\star$ & $\star$ & $\star$ & $0$ & $0$ & $\star$ 
& $0$ & $0$ \\
$T_i$  & $\star$ & $\eta_i(j)$ & $\star$ & $0$ & $\eta_i(0)$ & $0$ & 
$0$ & $\eta_i(-1)$ & $0$ \\
$T_N$  & $\star$ & $\star$ & $\star$ & $0$ & $0$ & $\star$ & $0$ 
& $0$ & $\star$ \bigstrut[b]\\
\end{block}
\end{blockarray}
\protect\caption{Likelihood ratio matrix layout (a simplified version of Figure 1 in \cite{VoB17})} \label{tab:layout}
\end{table*}

An entry in the first 
column block is the likelihood of a track having survived and being observed by 
that measurement, divided by the density of the measurement as clutter. 
The second column block is diagonal, and an entry on it is the probability of a track 
having survived but being mis-detected. The third column block is also diagonal, 
and an entry on it is the probability of a track having died. We take the negative 
log of the likelihood ratio matrix to get a cost matrix. A valid data association is 
defined by an assignment of the matrix such that each row has {\em exactly one} entry 
selected, and each column has {\em at most one} entry selected. The sum of the 
selected entries defines the cost of the association, the smaller the cost, the 
higher the likelihood.  

The best assignment can be found by using the Munkres algorithm 
\cite{MunkresJ57_algorithms}, while the best $K$ assignments  can be enumerated 
by using the Murty's algorithm \cite{MurtyK68}. Both have modern, faster 
versions; see \cite{LuQ18_murty} and the references therein. Since all current 
hypotheses perform this operation, and the union of their children constitute 
the next generation of hypotheses, a suboptimal but parallelizable selection 
scheme is to allocate, a priori, fixed number of children for each hypothesis, 
e.g., in proportion to its prior weight. The scheme is suboptimal because it 
may turn out that some child of a high-weight parent has a smaller weight than 
some would-be child of a lower-weight parent if the latter were given a larger 
allocation. 

If we implement the Murty's algorithm (or its variant) in the style of an {\tt 
iterator}, i.e., with methods such as {\tt has\_next()} and {\tt get\_next()}, 
then the optimal selection scheme\footnote{This idea was first proposed to the 
author by Peter Kingston.}  can be defined as follows in a round robin fashion:
\begin{enumerate}
 \item Let each hypothesis produce its best child. Put these in a selection 
buffer.
 \item Copy the best out of the buffer, and replace the content in this spot with the next 
best child from the same parent.
 \item Repeat until all top $K$ hypotheses have been obtained, or until no more 
children are available.
\end{enumerate}

Our observation has been that, with the Merge/Split algorithm, we typically do 
not need a large value for $K$. This will be illustrated in Section~\ref{sec:simu}.

\subsection{The Merge/Split algorithm}
Given coarse gating functions, incoming measurements together with existing tracks may form independent clusters. The end result of merging and splitting is to feed a factor only with its gated measurements, and the update of all the factors can be carried out in parallel. The factor falls into the following categories:
\begin{itemize}
 \item It is a newly created one, from a cluster of measurements that do not gate with any existing track.
 \item It is an existing one, or a merged one from existing ones that get coupled by new measurements.
 \item It is split off of an existing one or a merged one, either as the gated part or the non-gated part.
\end{itemize}

Details will be presented in the following.

\subsubsection{Clustering of measurements and tracks}
Upon the arrival of a new batch of measurements, the first step is to obtain 
clusters of measurements and tracks, such that entities from different clusters 
do not gate. Since we use the efficient algorithm presented in \cite{DezertJ93}, we construct an ``adjacency matrix'' as illustrated in Table~\ref{tab:mt},
\begin{table}[!htb]
\begin{center}
 \begin{tabular}{|c|c|c|c|c|c|}
 \hline
 & $z_1$ & $z_2$ & $z_3$ & $T_1$ & $T_2$ \\
 \hline 
 $z_1$ & 1 & * & * & * & * \\
 \hline
 $z_2$ & * & 1 & * & * & * \\
 \hline
 $z_3$ & * & * & 1 * & * & * \\
 \hline 
 \end{tabular}
 \end{center}
\caption{Illustration of an adjacency matrix between measurements and tracks.} \label{tab:mt}
\end{table}
where both the measurements $z_i$ and the tracks $T_j$ are actually represented by their respective IDs, and the Boolean values denoted by the stars are calculated through the coarse gating function: Two measurements gate if there exists a track state that gates with both, and a measurement gates with a track ID if it gates with any track with such an ID in any hypothesis in any factor.

Row merging operations on the adjacency matrix produces clusters of the following three types:
\begin{enumerate}[i)]
 \item It is a singleton track ID without any gated measurement. This is ignored for the time being; negative information update will be considered collectively later.
 \item It is a cluster of measurements without any gated track. This spawns a new factor and the update will be performed later together with other factors after they get their gated measurements.
 \item It is a cluster with both measurements and gated tracks. We will perform a secondary clustering operation to understand from which factors the tracks come from, and perform merging and splitting when applicable.
\end{enumerate}

\subsubsection{Super-clustering of clusters and factors}
Let $\{c_i\}$ denote the set of clusters obtained in the previous round, with each containing both measurements and tracks, and $\{f_j\}$ the set of existing factors. We perform yet another clustering operation on the ``adjacency matrix'' illustrated in Table~\ref{tab:cf}, where an entry is 1 if and only if there is a common Track ID that appears both in any track in the cluster and in any track in any hypo in the factor.
\begin{table}[!htb]
\begin{center}
 \begin{tabular}{|c|c|c|c|}
 \hline
 & $f_1$ & $f_2$ & $f_3$ \\
 \hline 
 $c_1$ & * & * & *  \\
 \hline
 $c_2$ & * & * & *  \\
 \hline
 \end{tabular}
 \end{center}
 
\caption{Illustration of an adjacency matrix between clusters and factors.} \label{tab:cf}
 
\end{table}
The operation produces what we call ``super-clusters'' that fall into the following categories:
\begin{enumerate}[i)]
 \item A factor that has no cluster, and therefore no measurements. It is queued up for negative information update.
 \item A set of factors with a set of measurements contained in the set of clusters. We will perform merging if there is more than one factor, and consider splitting if it is possible.
\end{enumerate}

\subsubsection{Merging} If a super-cluster has more than one factor, which are coupled by the new measurements, then they should be merged in the fashion of (\ref{eqn:merging}). Since we are operating under computational constraints, truncation of hypotheses is also considered in this step, which means that as we form the merged hypotheses, we want them to be sorted by weights, e.g., the first is the best from Factor 1 with the best from Factor 2, the second is the second best from Factor 1 with the best from Factor 2, and so on and so forth. Because of the ``layered'' structure of this problem, selection is performed using the efficient $K$-min-sum algorithm presented in the Appendix.

\subsubsection{Splitting} 
\begin{figure}[!htb]
 \centering
 \includegraphics[width=\columnwidth]{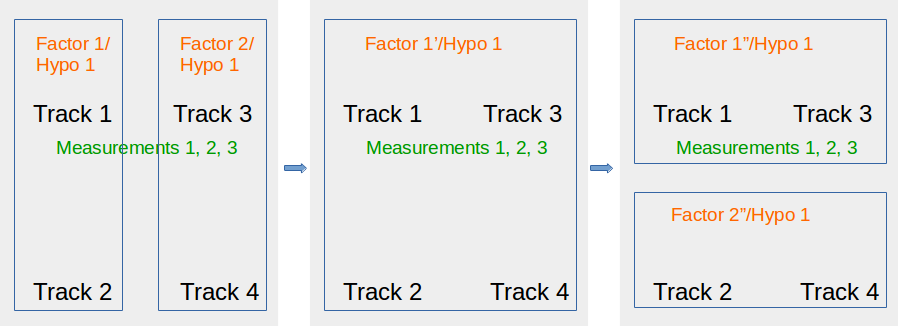}
 \caption{Illustration of the splitting step: Two hypos from two factors contain tracks that gate with a set of measurements (left panel), and therefore the two factors have to be merged. However, in the merged factor, only Tracks 1 and 3 gate with the measurements (middle panel), and therefore this factor may be split into two if independence holds approximately (right panel).} \label{fig:ms}
\end{figure}

The opportunity of splitting was possibly created after the measurement update at the {\em last} step. However, performing splitting at that time may turn out to be premature, because the smaller factors may get coupled by the measurements at the {\em current} time, and would have to be merged again. Therefore we try out splitting only after we have super-clustered the factors as described above. Now, not all tracks in a factor, whether existing or merged, necessarily gate with the coupling measurements, and we will determine whether we can split the factor along the gated/not-gated line, as illustrated in Figure~\ref{fig:ms}, where Factors 1 and 2 first merge to become Factor $1'$, and then possibly split into Factors $1''$ and $2''$.

After merging now-identical hypotheses as described in Section~\ref{sec:indexing}, and adopting the notations used in (\ref{eqn:LR}), we can identify sub-hypotheses from tracks that gate with the measurements as ``the left set,'' and the other, ``the right set,'' and pad with weight 0.0 for missing hypotheses. Thus we can construct the joint probability table as illustrated in Table~\ref{tab:ij}. Note that if we have not performed hypothesis management that combines duplicate hypotheses, then there may be repeated entries for the same $(i, j)$, in which case we simply add up their weights.
 
 \begin{table}[!htb]
 \begin{center}
\begin{NiceTabular}{cccc}[hvlines]
   \diagbox{$i$}{$j$}     & ...  &  $\left(\_, h_R^{(j)}\right)$  & ... \\ 
           ...             & ...  &  ...                             & ... \\ 
 $\left(h_L^{(i)}, \_\right)$ & ...  &          $w^{(ij)}$           & ... \\ 
           ...             & ...  &   ...                            &  ...
\end{NiceTabular}
\end{center}

\caption{Constructing the probability table $P(i, j)$, where weight is entered for a hypothesis  that shares a common ``left'' subset  across the row , and a common ``right'' subset across the column.} \label{tab:ij}
\end{table}
 
Now we can check whether independence can be considered to hold according to the reconstruction error defined by (\ref{eqn:eps}). If it can, then we split according to (\ref{eqn:split}) and create the left factor paired with the measurement set, and the right factor paired with the empty set. If independence does not hold, then we keep the original factor paired with the measurement set.

This concludes the preparation such that factors, both existing and newly created, do not interact with each other and are assigned their own gated measurements. Update can be performed in parallel, including negative information update when the measurement set is empty. Hypotheses with very small weights are discarded. A factor with only the empty hypothesis left in it is deleted.

\section{Some illustrations} \label{sec:simu}
We illustrate the Merge/Split algorithm with some simulations. The main point is to show that with this new approach, we do not necessarily need {\em many} hypotheses in order to track reasonably well.

\subsection{Simulation setup}
The fundamental units are meters and seconds in the following. A position sensor observes a rectangular region of size $600 \times 600$ with a Poisson distributed clutter whose uniform density is $\kappa = 10^{-5}$. The measurement covariance is $R = {\rm diag}([0.01, 0.01])$, and the probability of detection is $P_d = 0.99$.  A total of 36 birth models\footnote{Placing a grid of birth models helps mitigating the ``exploding symmetry'' problem described in Example 2 of \cite{GarciaFernandezA20_multiple}.} are placed uniformly inside this region, each with a birth probability $P_b = 0.02$, and a Gaussian distribution over position and velocity with a covariance $P_0 = {\rm diag}([400, 400, 25, 25])$. All targets born follow a near constant velocity motion model, with (continuous time formulated) process noise intensity in each direction $q = [0.01, 0.01]$ and a probability of survival $P_s \approx 1.0$.

Each Monte Carlo run draws samples from these distributions, and the measurements are passed to the tracker. The exact parameters are also assumed known to the tracker in this simulation; in practice they have to come from modeling and tuning. 

The factored $\delta$-GLMB tracker has some coarse gating parameters; here we simply define that two positions can gate only if their distance is less than 60. The number of factors are driven by data. However, each factor is given a budget of at most 10 hypotheses, both in merging, and in generating children. This is merely to illustrate what can be achieved with a factored approach, since the number 10 is significantly smaller than what is commonly assumed about $\delta$-GLMB.

The maximum reconstruction error $\epsilon$ for considering that independence holds is chosen as 0.01. 

\subsection{Simulation results} 
\subsubsection{The hypothesis ``tree''} \label{sec:tree} 
\begin{figure*}
 \centering
 %\fbox{
 \includegraphics[width=0.99\textwidth]{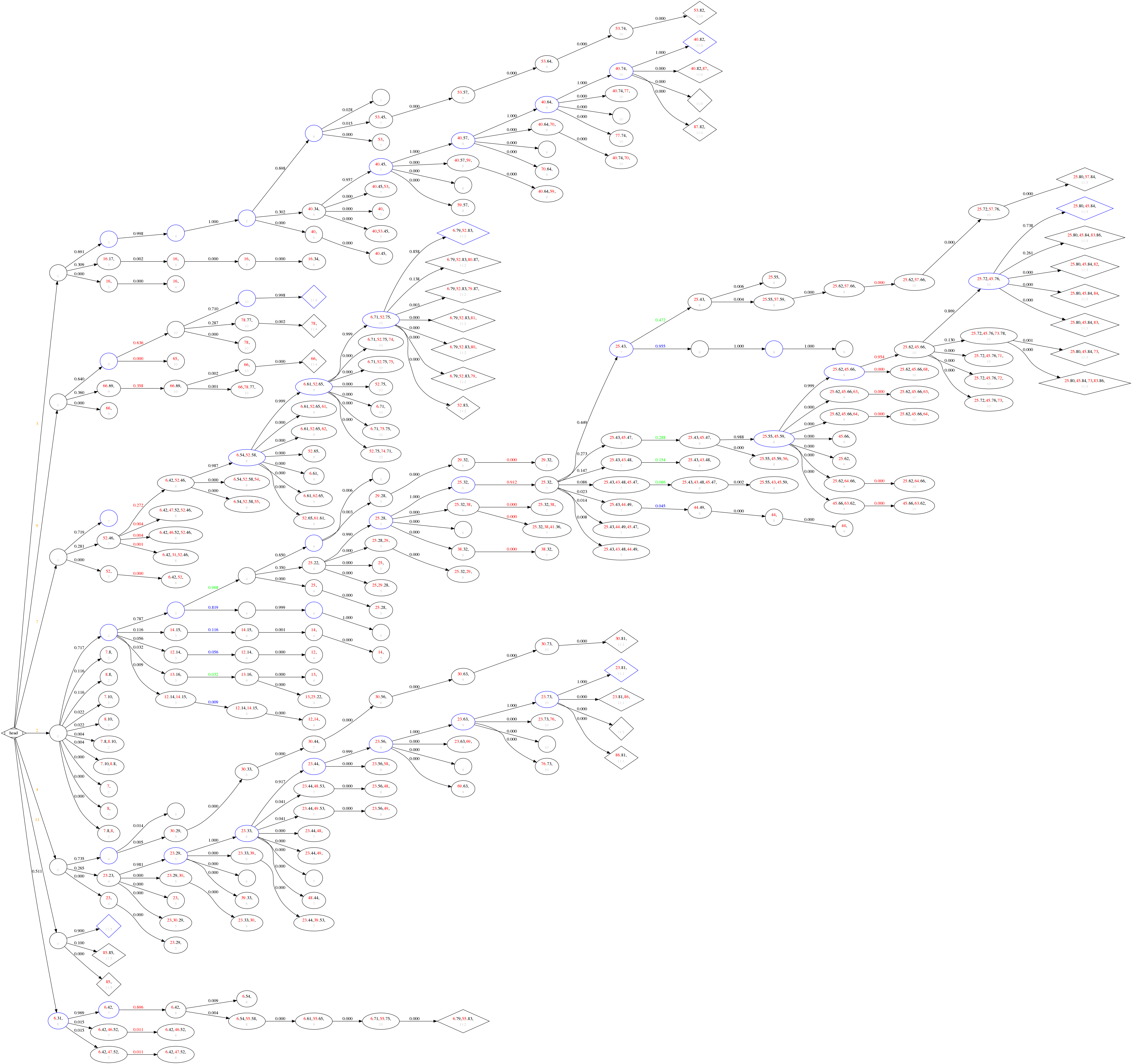} %,trim=0 1500 0 800,clip
%}
 \caption{History of hypotheses after the 11-th update. Explanations are given in Section~\ref{sec:tree}.
}
\label{fig:tree}
\end{figure*}

For real time execution, only the current hypotheses need to be kept in memory. In an offline debugging mode, we can also keep the pedigree in memory. To visualize a moving window of recent results, we start with the current hypotheses, trace their parents 10 generations up, and assign these ``great grandparents'' to a nominal ``root node'' so that we can render the structure as a tree for easy visualization. For merging and splitting, there is no single parent relationship. In such cases, we randomly choose one to be the parent, with a preference to the parent that shares a common track ID with the child. An example of such a hypothesis tree is shown in Figure~\ref{fig:tree}. 

The diagram shows the following:
\begin{itemize}
 \item Each ellipse or diamond represents a hypothesis, with the latter being for the current frame of update. The rendering of the tree with the nominal root node called ``head'' is such that hypotheses with a shorter history are located more to the left, as is the case in the lower left corner. 
 \item Each hypothesis shows a listing of comma delimited pairs such as {\tt {\textcolor{red}{12}}.34}, where the red number before the decimal point is the Track ID, while the black number is the Measurement ID and may be absent for missed detections. The weight of the hypothesis is shown on the incoming arrow.
 \item Each hypothesis also shows on the second line a gray number that is the Frame Number of the measurement set for its update. For the diamond, this number is followed by a colon and another number, which is the sequence number of the factor that contains this hypothesis. 
 \item The birth of a hypothesis is also shown with an arrow from the ``head'' node, and an orange integer on it denoting the frame number. 
 \item A simple algorithm is used to obtain an estimate of the tracks at each time: For each factor, the hypothesis with the largest weight defines the tracks. Such hypotheses are colored with a blue border.
 \item When merging happens, the weights are shown in the color red.
 \item When splitting happens, those hypotheses that gate with the measurements have their weights shown in green, while those that don't, in blue. Because of the many-to-many relationships between the ``before'' and ``after'' nodes, the chosen parent-child relationship is only for simple viewing as a tree. In other words, we have to aggregate all the green and blue weights to understand the underlying operation. 
\end{itemize}

\subsubsection{The track estimates} \label{sec:estimates} 
\begin{figure*}[!htb]
 \centering
 %\fbox{
 \includegraphics[width=0.99\textwidth,trim=0 70 15 85,clip]
 {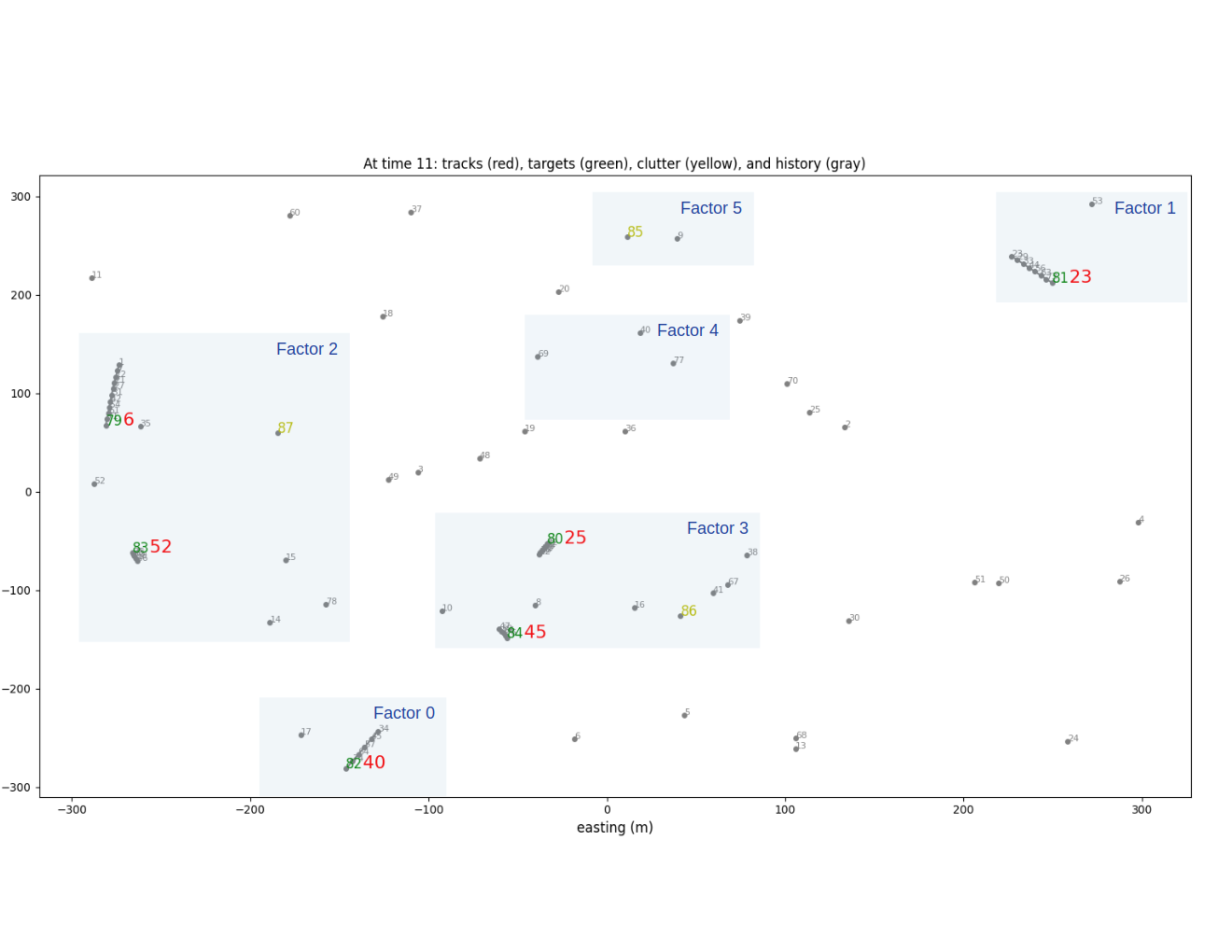} 
 %}
 \caption{Track estimates and measurements with history. Explanations are given in Section~\ref{sec:estimates}.} \label{fig:f11}
 
\end{figure*}

We also show in Figure~\ref{fig:f11} the track estimates after the 11-th frame of measurements, together with the current measurements and the history of measurements. 

More specifically,
\begin{itemize}
 \item A red number for Track ID is placed next to the mean of the position estimate of the track. The velocity estimate is not shown for simplicity of visualization.
 \item A measurement ID is placed next to the measured position. There can be three colors: The green indicates that the current measurement is generated from the true target, the orange indicates that the current measurement is sampled from the clutter distribution, while the gray, in a smaller font, indicates that the measurement is from previous frames. 
 \item For better visualization of correspondence, the factors shown in the diamonds in Figure~\ref{fig:tree} are marked in Figure~\ref{fig:f11}. However, a shaded rectangle does \emph{not} necessarily equal to the support of the factor; the latter is naturally and implicitly defined by the coarse gating functions of its constituent tracks.  
\end{itemize}

\section{Related work} \label{sec:related}
The algorithm presented in this paper is intimately related to various approaches in the literature, and we give a brief overview in this section. We apologize in advance if we miss any major contributions.

\subsection{The N-scan in Track Oriented MHT}
In TO-MHT \cite{MorefieldC77}, a new set of leaf nodes are created by considering all possible extensions of the current ones with a newly arrived frame of measurements, and any subset of the leaf nodes that are compatible (i.e., with no two nodes explaining the same track or the same measurement) forms a global hypothesis. Usually only the best global hypothesis is obtained by solving an Integer Linear Programming problem (minimizing the total cost which is the sum of the negative log likelihoods of the selected nodes, while satisfying the compatibility constraints), but the second, third, ... best can also be obtained by solving a modified ILP \cite{FortunatoE07}. 

So if there is no limit on time and memory and therefore no truncation, then both $\delta$-GLMB and TO-MHT produce the same set of hypotheses if they are explicitly computed. In reality, the explosion, of hypotheses in the former and of leaf nodes in the latter, has to be dealt with. The former truncates by taking only a subset of the top weighted hypotheses. The latter truncates not only by focusing just on the best global hypothesis, but also by pruning based on N-scan, which means that alternative explanations of data are kept alive in the hypothesis trees with the most recent N scans of measurements, but earlier decisions are made and committed according to the current best global hypothesis.

The result of this pruning is: No track has two leaf nodes with an identical measurement history that starts at a time earlier than the N-scan time, because there cannot still be branching at their common ancestor that is earlier than the N-scan time. In TO-MHT, leaf nodes are candidates that go in a global hypothesis when selected. In $\delta$-GLMB, ``track nodes'' are already constructed in all the enumerated hypotheses, so as time progresses we use a moving window of measurement IDs to identify those track nodes that have become ``identical,'' in order to merge duplicate hypotheses. We believe that this second level of truncation/pruning, achieved by the Merge/Split algorithm, is also needed for $\delta$-GLMB to be practically viable.

\subsection{Hypothesis management in Hypothesis-Oriented MHT}
In \cite{ReidD79}, hypothesis merging is presented in a similar way to our merging equation  (\ref{eqn:merging}). However, hypothesis splitting is not as rigorously presented. It seems that a large hypothesis becomes smaller only when a singleton can be split off from it, which happens with a confirmed track or a uniquely explained measurement, after weights below a threshold are discarded and pruning is performed. In our experience, singleton factors do happen, but often in a more general form with three hypotheses. To use the notations of Figure~\ref{fig:tree}, they are: $(T_i, M_j)$, which means that the track is observed by the measurement; $(T_i, )$, which means that the track is not detected in this frame and the measurement is clutter; and $()$, which means that the track has died and the measurement is clutter. In our framework, none of the above three hypothesis weights has to be small enough to be discarded in order for factoring to happen. We also note that a downside of always factoring out the singletons is that they may have to be merged again with the arrival of new measurements.

A splitting and merging algorithm for HO-MHT is extensively described in \cite{RoyJ97_efficient}. It is akin in spirit to our algorithm in this paper. However, the representation of hypothesis trees, the discussions around ``track family'' and ``asymmetry,'' and the concept of independence are more opaque than offered by a modern day treatment such as $\delta$-GLMB. More importantly, the reasoning in \cite{RoyJ97_efficient} seems to be ``Boolean'' rather than ``probabilistic'': It makes statements to the effect that if Track 1 is updated by Measurement 1, then it \emph{cannot} be updated by Measurement 2 in the same frame of input data. We know that in the RFS framework where the objective is to obtain a posterior density, rather than just the best estimate, both do happen, only with different probabilities, or weights. Thus in \cite{RoyJ97_efficient}, splitting seems to be dictated by pruning, without considering the effect of weights in the fashion of (\ref{eqn:eps}).

\subsection{Label-Partitioned GLMB}
Our approach was inspired by an early preprint on {\tt arxiv.org} that eventually became the journal publication of \cite{BeardM20_solution}. There is no emphasis on localized tracks in \cite{BeardM20_solution}, and the partitioning is distinguished by labels alone, hence ``Label-Partitioned GLMB.'' The same overall objective of exploiting statistical independence between labels and measurements across different groups is achieved by using two parameters: One is the probability $P_G$ that controls how stringent gating is, and the other is the maximum size $L_{\rm max}$ of any partition. Gating determines connected components that are candidates for partitions. If the size is too large, then $P_G$ is reduced and the process is repeated until the size constraint is satisfied. The {\em hope} is that the statistical dependence between different groups thus obtained is still negligible. In contrast, we control directly the tolerance of dependence through the reconstruction error $\epsilon$ in (\ref{eqn:eps}), and treat the coarse gating function as a given. This may result in a large factor when the intrinsic confusion is high. 

There is a subtle point about marginalization: The formula given by Proposition 1 in \cite{BeardM20_solution} has the same indexing variable $c$ in both the joint density and the marginalized density, and ``collapsing of terms'' (summing of weights) happens only \emph{within} the same $c$. For this to have a practical utility, the GLMB filter has to be implemented with quite some sophistication such that the set for $c$ is small while under each $c$ are more than one hypotheses. In the commonly adopted, straightforward implementation, each $c$ simply indexes \emph{one} hypothesis. But collapsing of terms can happen \emph{across} $c$, when the relevant kinematics densities $p(\cdot)$ are \emph{almost} the same if not \emph{exactly} the same. We identify such terms by indexing $p(\cdot)$ with a moving window of the incorporated measurement IDs, so that a GLMB with say 100 terms may be factored into two, each having only 10 terms instead of 100.

\subsection{Labeled Multi-Bernoulli filters}
The very popular Labeled Multi-Bernoulli (LMB) filter is used to 
approximate a $\delta$-GLMB filter in \cite{ReuterS14_labeled} and \cite{VoB14_towards}; it is 
effectively factorization down to single-track factors, and therefore can be 
considered as a special case in our framework. To perform measurement update, 
$\delta$-GLMB is reconstructed (losslessly, before truncation) from LMB, but to 
condense $\delta$-GLMB into LMB, a large approximation error {\em may} occur if 
independence is far from being valid. An adaptive scheme is proposed in 
\cite{DanzerA16_adaptive}  to judiciously switch between LMB and $\delta$-GLMB, taking 
advantage of the parsimony of the former, and the ``high resolution'' of the 
latter. The switching criteria include the Kullback-Leibler Criterion and the 
Entropy Criterion. Our Merge/Split algorithm is also an adaptive scheme that 
can switch between the ``singleton'' LMB, the ``full'' $\delta$-GLMB, and 
everything else in between, and the switching criterion, which is extremely light weight computationally, is simply based on 
comparing the joint probability reconstruction error with a predefined 
tolerance, in the fashion described by (\ref{eqn:eps}).

\subsection{Marginalized $\delta$-GLMB}
The Marginalized $\delta$-GLMB proposed in \cite{FantacciC16_scalable} first 
performs marginalization over the entire association history, then further 
combines the hypotheses through the use of multimodal densities. The first step 
can be considered as a special case of moving window size $N=1$ in our framework; we do not 
perform the second step.

\subsection{Correlation Analysis }
The paper \cite{YiW16_enhanced} considers a more general problem of approximating a Labeled Multi-Object (LMO) density with another density, possibly in factored form. The approach is to consider correlations between marginalized densities for each label, called basic components (BCs), and partition an RFS into groups such that BCs across different groups are uncorrelated. Then the factored form can be obtained through marginalization. 

Although such factorization based on correlation analysis has theoretical appeal, it is unclear whether it can provide a practically viable algorithm. Apart from the issue that pairwise correlation scales quadratically with the number of targets and is probably not all needed, marginalization is only given as a {\em formal} sum for $\delta$-GLMB, which means that it does not ``collapse terms'' to yield a simpler representation. We belabor this point as follows: Unlike marginalization over a discrete probability table that collapses terms by adding them as real numbers, marginalization of $\delta$-GLMB involves both the weights and the kinematic densities. Although formally we can ``combine'' $\alpha \, p_1(x)$ with $ \beta \, p_2(x)$ to get $(\alpha + \beta) \, \bar{p}(x)$ by defining 
\[\bar{p}(x) \triangleq \frac{\alpha}{\alpha + \beta} \, p_1(x) + \frac{\beta}{\alpha + \beta} \, p_2(x),
 \]
thus seemingly reducing two terms to one, this merely trades the complexity in multi-hypothesis for the complexity in multi-mode. It is only when $p_1(x) \approx p_2(x) \approx 
\bar{p}(x)$ that real savings can be achieved by using one density and summing up the weights.

\section{Conclusions} \label{sec:con}
In this paper we have presented a Merge/Split algorithm to efficiently manage 
hypotheses in the framework of multitarget tracking using $\delta$-GLMB 
filters. A factored representation of the posterior density is maintained, by 
merging factors that are coupled by new measurements, and by splitting the 
merged factor if independence condition is met. Through indexing kinematic densities by the measurement IDs that have updated them, in a moving window, marginalization of the hypothesis densities is transformed into a discrete problem and can be efficiently and effectively achieved, so that independence within a given tolerance can be checked. 

Under a computation resource constraint, a filtering scheme based on hypothesis enumeration and truncation by weight is prone to degeneracy that would fail to initialize new tracks, and factorization mitigates this problem. This is illustrated with a simple simulation example where a surprisingly few number of hypotheses can achieve adequate tracking results, with merging and splitting that is automatically driven by data.

\section*{Acknowledgment}
The author would like to thank his current and former colleagues, in particular 
Drs. Giancarlo Baldan, Yu Pan, and Peter Kingston, for their valuable inputs 
and help; Dr. Angel Garcia-Fernandez for insightful discussions; and reviewers of earlier versions of this paper for constructive suggestions.

\bibliographystyle{IEEEtran}
\bibliography{MyBSTcontrol,rfs}

\section*{Appendix: $K$-min-sum algorithm}
 Let $\{A_i, i = 1, \ldots, N\}$ be a set of arrays of real numbers. A {\em 
selection} $S_N \triangleq \{ s_i, i = 1, \ldots, N\}$ is a sequence of indices 
into the corresponding arrays, and with an abuse of notation we denote the sum 
of the selected numbers $\sum_{i=1}^N A_i[s_i]$ also by $S_N$. We seek the top 
$K$ such selections with the smallest sums: $T^{K}_N \triangleq \{S_N^{(i)}, i 
= 1, \ldots, K, S_N^{(i)} \le S_N^{(j)} \, \forall j > K \}$, where the parenthesized superscript is merely an index. The solution can 
be obtained recursively as follows.
  \begin{enumerate}
   \item For $N = 1$, $T^{K}_1$ is the $K$ smallest elements of the array $A_1$.
   \item Suppose we have obtained $T^{K}_{N-1}$. 
   \item Enumerate candidate selections by extending each in $T^{K}_{N-1}$ with 
an element in $A_N$. Take the top $K$ smallest, and that gives $T^{K}_N$. The 
brute force enumeration can be replaced by a more efficient procedure of 
``popping'' two sorted queues, but the gain is not significant for small values 
of $K$.
  \end{enumerate}
We prove the correctness of the algorithm by contradiction. For simplicity we 
assume that there is no tie in sum comparison. Suppose $\bar{S}_N \triangleq 
\{\bar{s}_1, \ldots, \bar{s}_N\}$ is one of the top $K$ selections but is 
missed by our algorithm. Then we reason as follows.
\begin{itemize}
 \item First, we conclude that $\bar{S}_{N-1} \triangleq \{\bar{s}_1, \ldots, 
\bar{s}_{N-1} \} \not\in T^{K}_{N-1}$, because otherwise, by Step 3 of the 
algorithm, $\bar{S}_N$ would be one of the enumerated candidates and should not have 
been missed.
 \item Second, the existence of $\bar{S}_{N-1} \not\in T^K_{N-1}$ means that 
the cardinality of $T^K_{N-1}$ is no less than $K$.
\item Third, let the smallest element in $A_N$ be $a$. Then every extension of  
$T^{K}_{N-1}$ by $a$ is smaller than $\bar{S}_N$. But there are already $K$ of the 
former, thus contradicting the assumption that $\bar{S}_N$ is in top $K$. $\blacksquare$
\end{itemize}

\end{document}